# Data-Dependent Early Completion of Dose Finding Trials for Drug-Combination

Masahiro Kojima[1,2]

[1]Biometrics Department, R&D Division, Kyowa Kirin Co., Ltd., Tokyo, Japan.

[2]Department of Statistical Science, School of Multidisciplinary Sciences, The Graduate University for Advanced Studies, Tokyo, Japan.

**Acknowledgments**: The author thanks Professor Masahiko Gosho and Associate Professor Hisashi Noma for their encouragement and helpful suggestions.

**Corresponding author**

Name: Masahiro Kojima

Address: Biometrics Department, R&D Division, Kyowa Kirin Co., Ltd.

Otemachi Financial City Grand Cube, 1-9-2 Otemachi, Chiyoda-ku, Tokyo, 100-004, Japan.

Tel: +81-3-5205-7200

FAX: +81- 3-5205-7182

Email: masahiro.kojima.tk@kyowakirin.com

**Running title**: Early Completion of Drug-Combination trials

**List of where and when the study has been presented in part elsewhere:** None

**Financial support**: None

**Conflict of interest disclosure statement**: None

**Context Summaries:**

We propose a data-dependent early completion of dose finding trials for drug-combination. The early completion is determined when a beta-binomial probability for dose retainment with the trial data and the number of remaining patients is high. This paper also proposes an early completion method that a dose retainment probability is adjusted by a bivariate isotonic regression. We demonstrate the early completion for a virtual trial. We evaluate the performance of early completion method through simulation studies with 12 scenarios. We confirmed the superior performance for our proposed early completion methods. We show the number of patients for determining early completion before a trial starts and a program code for calculating dose retainment probability in this paper.**Word count**: 2998

**The number of figures**: 3

**The number of tables**: 3

# Abstract

**Purpose:** Model-assisted designs for drug combination trials have been proposed as novel designs with simple and superior. The model-assisted designs require to set a sample size in advance, and the trial cannot complete until the number of patients treated reaches the sample size. We propose a data-dependent early completion of dose finding trials for drug-combination.

**Methods**: The early completion is determined when a beta-binomial probability for dose retainment with the trial data and the number of remaining patients is high. This paper also proposes an early completion method that a dose retainment probability is adjusted by a bivariate isotonic regression. We demonstrate the early completion for a virtual trial. We evaluate the performance of early completion method through simulation studies with 12 scenarios.

**Results**: From simulation studies, the percentage of early completion was averagely determined about 70% and the number of patients treated reduced by 25% from planned sample size. The percentage of a correct maximum tolerated dose-combination selection for early completion designs showed little change from the non-early completion designs by an average of about 3%.

**Conclusion**: We confirmed the superior performance for our proposed early completion methods. We showed the number of patients for determining early completion before the trial starts and a program code for calculating dose retainment probability in this paper.

# Introduction

We consider a dose-finding for drug-combination trial that aims to identify a maximum tolerated dose-combination (MTD). The MTD is a dose-combination whose dose limiting toxicity (DLT) rate is closest to a target toxicity level (TTL). To identify the MTD, dose adjustments are repeatedly needed for experience multiple doses. A design for the dose adjustments is classically a 3+3 design[1]. The 3+3 design is simple to operate for the dose adjustment based on a simple algorithm, but the performance of correct MTD selection is poor. Hence, designs based on a statistical model have been proposed[2,3,4,5,6,7,8,9,10,11,12,13,14,15]. However, the model-based designs are rarely used in actual clinical trials because the designs require complex assumptions[16]. Therefore, a Bayesian optimal interval (BOIN)[17,18,19], Keyboard[20,21], and extended model-assisted designs[22,23,24,25,26,27,28,29,30,31,32,33] have been proposed as simple and good performing. The BOIN and Keyboard designs can prepare a dose-assignment decision table based on a simple statistical model before the trial starts. The BOIN and Keyboard designs require to set a sample size in advance, and the trial cannot complete until the number of patients treated reaches the sample size. Although the BOIN and Keyboard designs have a stopping rule to terminate the trial if the number of patients treated exceeds the predetermined number, there is no statistical evidence for the number.

Kojima[28,29,31] proposed an early completion method for model-assisted designs, but Kojima did not discuss an applicability of early completion for dose-combination trials.

In this paper, we propose a data-dependent early completion of dose finding trials for drug-combination. The early completion is determined when a beta-binomial probability for dose retainment with the trial data and the number of remaining patients is high. This paper also proposes an early completion method that a dose retainment probability is adjusted by a bivariate isotonic regression. We demonstrate the early completion for a virtual trial. We evaluate the performance of early completion method through simulation studies with 12 scenarios.

The paper is structured as follows: Section 2 presents early completion methods for dose-combination trials. In addition, we demonstrate the early completion for a virtual trial and present a simulation configuration to evaluate a performance for the early completion method. Section 3 presents the simulation results. Section 4 discusses our proposed designs and the results.

## Methods

We consider a drug-combination trial with the following: a sample size of $N$, $J$ dose levels of drug A and $K$ dose levels of drug B, a current combined-dose level $(j,k)$ of $d_{j,k}$, the total number of patients treated at $d_{j,k}$ is $n_{j,k}$, the total number of DLTs at the current dose $d_{j,k}$ is $m_{j,k}$, the observed DLT rate at the current dose is $\hat{p}_{j,k} = m_{j,k}/n_{j,k}$, the TTL is $\phi$, and a prior distribution of each dose-combination is Beta(1,1)(=Uniform(0,1)).

Bayesian optimal interval (BOIN) design conducts a dose-assignment compared $\hat{p}_{j,k}$ and a proper dosing interval $I_{pro,BOIN} = (\lambda_e(\phi), \lambda_d(\phi))$ for the DLT rate around a target DLT level. $\lambda_e(\phi)$ is the maximum value at which a dose escalation is determined and $\lambda_d(\phi)$ is the minimum value at which a dose de-escalation is determined. The boundaries of the proper dosing interval are calculated by minimizing the misjudgment of dose adjustment. For example, when $\phi = 0.3$, $I_{pro,BOIN} = (0.236, 0.358)$. Keyboard design conducts a dose-assignment compared an interval probability of a proper dosing interval $I_{pro,Key} = (\phi - 0.05, \phi + 0.05)$ called a target key and other interval probabilities. When we do not distinguish $I_{pro,BOIN}$ and $I_{pro,Key}$, we write just $I_{pro}$.

For drug-combination trials, if the dose for next cohort also retains the current dose, the dose adjustment is simple. On the other hand, the problems with combination trials are that there are many ways to the dose escalation and de-escalation. It would be nice to be able to try all combinations without safety issues, but a strategic dose selection is necessary to identify MTD in a limited sample size. Figure 1 shows the strategy for dose-assignment. The two candidates for dose escalation from dose combination $d_{1,1}$ were the blue dotted line for $d_{1,2}$ and the blue solid line for $d_{2,1}$, and the dose was escalated to dose $d_{2,1}$ whose DLT rate was closer to the TTL. Subsequent dose escalations were selected in the same manner. If a dose de-escalation is determined after the administration of dose combination $d_{3,2}$, the dose-combination de-escalates to $d_{3,1}$ or $d_{2,3}$ whose DLT rate is closer to the TTL. Pan et al.[19] proposed algorithms to escalate or de-escalate the doses of two drugs simultaneously, but this paper deals with the method to increase or decrease only one dose of two drugs. We summarize the algorithm for dose escalation/de-escalation in the dose-combination trial below.

**[Algorithm of dose escalation/de-escalation]**
- When the dose combination for next cohort is determined as an escalation after $d_{j,k}$ is administered, escalate to a dose combination which has the highest interval posterior probability of the interval $I_{pro}$ for either $d_{j+1,k}$ or $d_{j,k+1}$. If the candidate combination for dose escalation has never been administered, the interval probability is calculated by using the prior distribution. If the interval probabilities are equal, choose a drug combination randomly.

- When the dose combination for next cohort is determined as a de-escalation after $d_{j,k}$ is administered, de-escalate to a dose combination which has the highest interval posterior probability of the interval $I_{pro}$ for either $d_{j+1,k}$ or $d_{j,k+1}$. If the candidate combination for dose escalation has never been administered, the interval probability is calculated using the prior distribution. If the interval probabilities are equal, choose a drug combination randomly.

If the drug combination administered is too toxic $(P(p_{j,k} > \phi) > 0.95)$, stop the administration of the current dose combination and all higher dose combinations in the dose-finding trial. After completion of the trial, the MTD is selected as the dose combination whose observed toxicity rate is closest to the TTL. Because of the small sample size of the dose-finding trial, the observed toxicity rates are adjusted by a bivariate isotonic regression[34] to give a monotonically increase for the toxicity rates, and then the MTD is selected. When there are multiple dose combinations that are closest to the TTL, if the observed toxicity rates of the multiple dose combinations are below the TTL, we select the highest dose combination as MTD. If the observed toxicity rates of the multiple dose combinations are over the TTL, we select the lowest dose combination as MTD. If there are simultaneously the multiple dose combination with higher and lower observed toxicity rates than the TTL, we select randomly the minimum dose combination in the dose combinations with higher observed toxicity and the maximum dose combination in the dose combinations with lower observed toxicity. The BOIN and Keyboard designs require to set a sample size in advance, and the trial cannot complete until the number of patients treated reaches the sample size. Although the BOIN and Keyboard designs have a stopping rule to terminate the trial if the number of patients treated exceeds the predetermined number, there is no statistical evidence for the number.

**Early completion method**

We propose a data-dependent early completion of dose finding trials for drug-combination. The early completion is determined when a beta-binomial probability for dose retainment with the trial data and the number of remaining patients is high. The BOIN and keyboard designs can summarize the number of patients treated for which a dose retainment is determined in a table and the dose retainment probability uses this number. We assume that $\mathcal{R}$ is the set of the number of patients for dose retainment for $n_{j,k} + l$. For example, $\mathcal{R}$ includes 3 and 4 patients, when the number of patients is 12 patients for the BOIN design. The dose retainment probability is calculated by the equation below,

$$\text{dose retainment probability} = \sum_{r \in \mathcal{R}} BetaBinom(r - m_{j,k}; l, m_{j,k} + 1, n_{j,k} - m_{j,k} + 1). \quad \text{①}$$

*BetaBinom* is the beta-binomial probability function that $r - m_{j,k}$ DLTs occur in the remaining $l$ patients when there are $m_{j,k}$ DLTs for $n_{j,k}$ patients. The trial is completed early when the dose retainment probability over a threshold. Kojima[29] recommends a threshold of 0.4 because the dose retainment interval for the DLT rate for BOIN design is a maximum of 0.4. In addition, because the MTD is selected by using the toxicity rate adjusted by a bivariate isotonic regression, we consider the early completion method using the toxicity rate adjusted by the bivariate isotonic regression. If the current dose-combination is the maximum dose-combination, $\mathcal{R}$ includes the number of patients for which a dose escalation is determined. Because the $\mathcal{R}$ includes not only the number of patients for dose retainment but also the number of patients for dose escalation, the dose retainment probability ① is divided by 2. The maximum dose-combination means the maximum dose of both two drugs, and if the planned maximum doses are changed by the stopping rule, the maximum doses are also changed. The minimum dose is calculated in the same way. We also propose the dose retainment probability which is calculated by replacing $m_{j,k}$ in the equation ① with $n_{j,k} \times$ *adjusted toxicity rate.*

We demonstrate the early completion for a virtual trial. From here, we call the normal dose retainment probability DRP and the dose retainment probability based on the bivariate isotonic regression DRP-I.

**Example trial**

We assume the dose-finding trial with the 3 dose levels of drug A and the 3 dose levels of drug B, the sample size of 30, the cohort size of 3, the TTL of 30, and the threshold for early completion of 0.4. We show the result of 24 patients treated to Table 2. Dose adjustments for up to 24 patients are shown in Supplemental Example. The current dose combination is $d_{2,2}$ and the safety evaluation has been completed. $\mathcal{R}$ for 15 patients (9 patients treated and 6 remaining patients) for the BOIN design is {4,5}. Because $\mathcal{R}$ of the Keyboard design is the same as one of the BOIN design, we consider an early completion without distinguishing between the designs.

The DRP is calculated below,

$$BetaBinom(4-3; 6, 3+1, 9-3+1) + BetaBinom(5-3; 6, 3+1, 9-3+1) = 0.493.$$

The DRP is over the threshold of 0.4. Hence, this trial halts early.

Next, we calculate the DRP-I. The observed DLT rates adjusted by the bivariate isotonic regression are

$$\begin{pmatrix} 0.000 & 0.000 & 0.000 \\ 0.167 & 0.335 & 0.664 \\ 0.000 & 0.664 & 0.000 \end{pmatrix},$$

the dose corresponding to each element is

$$\begin{pmatrix} d_{1,1} & d_{1,2} & d_{1,3} \\ d_{2,1} & d_{2,2} & d_{2,3} \\ d_{3,1} & d_{3,2} & d_{3,3} \end{pmatrix}.$$

The adjusted $m_{2,2}$ is 3.015. DRP-I is 0.491. The DRP-I is over the threshold of 0.4. Hence, this trial halts early. This trial can complete without dosing six patients by the early completion method. The six fewer patients mean a six-month reduction in trial duration, for example, if the safety evaluation period is one month and the enrollment of patient takes one month. We described the sample R program code to calculate the dose retainment probabilities in Supplemental material.

**Simulation study**

**Simulation settings.** We demonstrated the performance of the early completion methods via a Monte Carlo simulation. We prepared 6 designs: BOIN, BOIN using the early completion method based on DRP (BOIN-EC), BOIN using the early completion method based onDRP-I (BOIN-ECI), Keyboard (Key), Keyboard using the early completion method based on DRP (Key-EC), Keyboard using the early completion method based on DRP-I (Key-ECI). The dose-combinations were assumed 12 scenarios including two sizes of dose matrices ($3 \times 4$ and $5 \times 6$). The detailed true DLT rates for each scenario were provided in Supplemental Table 1. The sample sizes were 45 for $3 \times 4$ drug combinations and 90 for $5 \times 6$ drug combinations. The cohort size was 3. The target toxicity level (TTL) was 30%. The threshold for early completion was 0.4. Sensitivity analyses also were performed by changing the sample size to 36 for $3 \times 4$ and 75 for $5 \times 6$ and the TTL to 20%, respectively. In addition, we conduct sensitivity analyses with changed to the threshold for early completion to 0.35 and 0.45. The simulation study was conducted in R. We evaluated each method using the following criteria.

**Evaluation criteria.**

1. Percentage for correct MTD selection (PCMS).

2. Percentage for doses selection lower correct MTD

3. Percentage for doses selection higher correct MTD

4. The average number of patients treated

5. Percent change of patients treated from planned sample size

6. Percentage for early completion

# Results

**Percentage for the correct MTDs selection (PCMS).** Figure 2 shows the results of PCMSs for 12 scenarios. The change in PCMSs of BOIN-EC from BOIN averaged -3.9%, with a minimum -0.9% at scenario 6 and a maximum of -9.9 at scenario 9. The change in PCMSs of BOIN-ECI from BOIN averaged -4.6%, with an increasing maximum of 4.0% at scenario 6 and a decreasing maximum of -8.6 at scenario 10. The change in PCMSs of Key-EC from Key averaged -2.3%, with an increasing maximum of 0.8% at scenario 1 and a decreasing maximum of -4.8 at scenario 2. The change in PCMSs of Key-ECI from Key averaged -1.7%, with an increasing maximum 1.4% at scenario 6 and a decreasing maximum of -3.8% at scenario 9. The change in keyboard design using the early completion method was smaller than the BOIN design.

**Percentage for doses selection lower correct MTD.** We show the difference (the result of early completion designs － the result of non-early completion design). For BOIN-EC, the average was 3.5%, with a minimum of 0.9% at scenario 1 and a maximum of 6.8% at scenario 9. For BOIN-ECI, the average was 7.0%, with a minimum 3.6% at scenario 12 and a maximum 13.4% at scenario 7. For Key-EC, the average was 1.2%, with a minimum of -0.7% at scenario 12 and a maximum of 2.5% at scenario 10. For Key-ECI, the average was 2.2%, with a minimum of 0.4% at scenario 12 and a maximum of 3.1% at scenario 3. The change of Key-EC was the smallest.

**Percentage for doses selection higher correct MTD.** We show the difference (the result of early completion designs − the result of non-early completion design). For BOIN-EC, the average was 0.7%, with a minimum of -2.3% at scenario 7 and a maximum of 3.7% at scenario 2. For BOIN-ECI, the average was -2.0%, with a minimum -6.5% at scenario 7 and a maximum 1.3% at scenario 9. For Key-EC, the average was 1.2%, with a minimum of -1.3% at scenario 7 and a maximum of 4.7% at scenario 2. For Key-ECI, the average was -0.3%, with a minimum of -2.3% at scenario 7 and a maximum of 1.9% at scenario 12. The change of BOIN-EC was the smallest.

**The average number of patients treated/ Percent change of patients treated from planned sample size.** For $3 \times 4$ drug combinations with the sample size of 45, the percent changes of patients treated from planned sample size for BOIN-EC and BOIN-ECI were averagely -24.6% and -25.1% (reduced about 11 patients), with a minimum of -14.2% and -14.0% (reduced about 6 patients) at scenario 8 and a maximum of -54.7% and -60.0% (reduced about 25 patients) at scenario 6. For Key-EC and Key-ECI, the percent changes were averagely -20.6% and -17.4% (reduced about 9 patients), with a minimum of -11.3% and -8.4% (reduced about 4 patients) at scenario 8 and a maximum of -48.2% and -47.6% (reduced about 23 patients) at scenario 6. For 5 × 6 drug combinations with the sample size of 90, the percent changes of patients treated from planned sample size for BOIN-EC and BOIN-ECI were averagely -31.6% and -29.8% (reduced about 27 patients), with a minimum of -28.1% and -26.2% (reduced about 24 patients) at scenario 11 and a maximum of -34.0% and -32.7% (reduced about 30 patients) at scenario 10. For Key-EC and Key-ECI, the percent changes were averagely -21.2% and -20.1% (reduced about 19 patients), with a minimum of -17.4% and -16.2% (reduced about 15 patients) at scenario 11 and a maximum of -24.0% and -22.9% (reduced about 21 patients) at scenario 10. On average, all changes were similar.

**Percentage for early completion.** For 3 × 4 drug combinations, the percentages for early completion of BOIN-EC and BOIN-ECI were averagely 69.6% and 64.2%, with a minimum of 53.4% and 45.9% at scenario 8 and a maximum of 94.0% and 94.1% at scenario 6. For Key-EC and Key-ECI, the percentages were averagely 64.7% to 57.1%, with a minimum of 47.9% and 39.0% at scenario 8 and a maximum of 92.2% and 90.9% at scenario 6. For 5 × 6 drug combinations, the percentages for early completion of BOIN-EC and BOIN-ECI were averagely 85.5% and 82.0%, with a minimum of 80.1% and 75.5% at scenario 11 and a maximum of 90.3% and 86.0% at scenario 10. For Key-EC and Key-ECI, the percentages were averagely 75.2% to 71.0%, with a minimum of 67.7% and 63.1% at scenario 11 and a maximum of 82.1% to 77.3% at scenario 10. BOIN-EC and Key-EC had the higher percentage of early completion on average.

## Discussion

We proposed the early completion methods for dose finding trials for drug-combination. The early completion is determined when a beta-binomial probability for dose retainment with the trial data and the number of remaining patients is high. We also proposed an early completion method that a dose retainment probability is adjusted by a bivariate isotonic regression.

We demonstrated the early completion for the virtual trial with $3 \times 3$ drug combinations and sample size of 36. By completing the trial early, we reduced 6 patients from the sample size. The six fewer patients mean a six-month reduction in trial duration, for example, if the safety evaluation period is one month and the enrollment of patient takes one month.

For the simulation study, the percentages for the correct MTDs selection (PCMSs) in all early completion designs were almost the same as compared non-early completion designs. The Keyboard designs with early completion methods were smaller the change in PMCSs from the keyboard design. The results of the sensitivity analyses with changed planned sample size or TTL were observed same trend. In comparison with the stopping rule proposed by Trialdesign.org, the overall PCMSs of the designs with the stopping rule were lower than our proposed method, and the PCMSs reduced by about 10% compared with non-early completion designs in scenarios 9-11 in Supplemental Table 9. We consider that the number of remaining patients should be considered for the early completion. By the application of early completion method, we confirmed the non-higher percentage of lower/higher doses selection than correct MTD compared with non-early completion designs. We confirmed that the early completion was averagely determined about 70% and the number of patients treated reduced averagely about 20-30%.

      We confirmed the superior performance for our proposed early completion methods. In specially, the keyboard with early completion was superior. Among the two early completion methods, ECI was slightly superior. However, we recommend EC because it is easier to implement. Our proposed method can be applied to dose-finding trial with more two drugs in the same way. We can show the number of patients for determining early completion before the trial starts and show in Table 3. This table can be also applied to the single drug and more two drugs dose-finding trial. We described the sample program code of early completion in supplemental material.

**Author's Contributions**

M. Kojima: Conception and design; development of methodology; acquisition of data (provided animals, acquired and managed patients, provided facilities, etc.): analysis and interpretation of data (e.g., statistical analysis, biostatistics, computational analysis); writing, review, and revision of the manuscript; administrative, technical, and material support (i.e., reporting and organizing data, constructing databases); and study supervision.

**References**

1. Simon R, Rubinstein L, Arbuck SG, Christian MC, Freidlin B, et al, Accelerated titration designs for phase I clinical trials in oncology, J. Natl. Cancer Inst. 1997;89:1138–1147.

2. O'Quigley J, Pepe M and Fisher L. Continual reassessment method: a practical design for phase 1 clinical trials in cancer. Biometrics 1990; 46: 33–48.

3. Thall PF, Millikan RE, Muller P, Lee SJ. Dose-finding with two agents in phase I oncology trials. Biometrics 2003; 59: 487–496.

4. Conaway MR, Dunbar S and Peddada SD. Designs for single- or multiple-agent phase I trials. Biometrics 2004; 60: 661–669.

5. Ivanova A and Wang K. A nonparametric approach to the design and analysis of two-dimensional dose-finding trials. Stat Med 2004; 23: 1861–1870.

6. Huang X, Biswas S, Oki Y, et al. A parallel phase I/II clinical trial design for combination therapies. Biometrics 2007; 63: 429–436.

7. Fan SK, Venook AP and Lu Y. Design issues in dose-finding phase I trials for combinations of two agents. J Biopharm Stat 2009; 19: 509–523.

8. Yin G and Yuan Y. Bayesian dose finding in oncology for drug combinations by copula regression. J R Stat Soc Ser C Appl Stat 2009; 61: 211–224.

9. Wages NA, Conaway MR and O'Quigley J. Dose-finding design for multi-drug combinations. Clin Trials 2011; 8: 380–389.

10. Lee BL and Fan SK. A two-dimensional search algorithm for dose-finding trials of two agents. J Biopharm Stat 2012; 22: 802–818.

11. Harrington JA, Wheeler GM, Sweeting MJ, et al. Adaptive designs for dual-agent phase I dose-escalation studies. Nat Rev Clin Oncol 2013; 10: 277–288.

12. Hirakawa A, Hamada C and Matsui S. A dose-finding approach based on shrunken predictive probability for combinations of two agents in phase I trials. Stat Med 2013; 32: 4515–4525.

13. Liu S and Ning J. A Bayesian dose-finding design for drug combination trials with delayed toxicities. Bayesian Anal 2013;8: 703–722.

14. Lam KC, Lin R, Yin G. Non-parametric overdose control for dose finding in drug combination trials. JRSS Appl Statist 2019;68: 1111–1130.

15. Razaee SZ, Cook-Wines G, Tighiouart M. A nonparametric Bayesian method for dose finding in drug combinations cancer trials. Stat Med 2022:1-22

16. Riviere KM, Le Tourneau C, Paoletti X, Dubois F, Zohar S. Designs of drug-combination phase I trials in oncology: a systematic review of the literature. Annals of Oncology

2015;26:669–674

17. Liu S, Yuan Y. Bayesian optimal interval designs for phase I clinical trials. Appl Statist 2015;64:507-523.

18. Yuan Y, Lee JJ, Hilsenbeck GS, Model-Assisted Designs for Early-Phase Clinical Trials: Simplicity Meets Superiority. JCO PO 2019;3;1-12

19. Pan H, Lin R, Zhou Y, Yuan Y. Keyboard design for phase I drug-combination trials. Contemporary Clinical Trials 2020;92:1-11

20. Yan F, Mandrekar JS, Yuan Y. Keyboard: A Novel Bayesian Toxicity Probability Interval Design for Phase I Clinical Trials. Clin Cancer Res 2017;23(15):3994-4003.

21. Lin R, Yin G. Bayesian optimal interval design for dose finding in drug-combination trials. Statistical Methods in Medical Research 2017;26(5):2155-2167

22. Takeda K, Taguri M, Morita S. BOIN-ET: Bayesian optimal interval design for dose finding based on both efficacy and toxicity outcomes. Pharma Stat 2018,17:383-395.

23. Yuan Y, Lin R, Li D, Nie L, Warren KE. Time-to-Event Bayesian Optimal Interval Design to Accelerate Phase I Trials. Clin Cancer Res 2018;24:4921-4930.

24. Mu R, Yuan Y, Xu J, Mandrekar SJ, Yin J. gBOIN: a unified model-assisted phase I trial design accounting for toxicity grades, and binary or continuous end points. JRSS Appl

Statist 2019;68: 289–308.

25. Lin R, Yuan Y. Time-to-event model-assisted designs for dose-finding trials with delayed toxicity. Biostatistics 2020;21:807–824.

26. Lin R, Zhou Y, Yan F, Li D, Yuan Y. BOIN12: Bayesian Optimal Interval Phase I/II Trial Design for Utility-Based Dose Finding in Immunotherapy and Targeted Therapies. JCO PO 2020;4:1393-1402

27. Takeda, K, Morita S, Taguri M. TITE-BOIN-ET: time-to-event Bayesian optimal interval design to accelerate dose-finding based on both efficacy and toxicity outcomes. Pharma Stat 2020;19:335–349.

28. Kojima M. Early completion of phase I cancer clinical trials with Bayesian optimal interval design. Stat Med 2021;40:3215-3226

29. Kojima M. Early completion of model-assisted designs for dose-finding trials. JCO Precis Oncol 2021: 5, 1449-1457

30. Kojima M. Adaptive design for identifying maximum tolerated dose early to accelerate dose-finding trial. Arxiv. 2021;Arxiv ID: 2110.02413.

31. Kojima M. Early completion based on multiple dosages to accelerate maximum tolerated dose-finding. Arxiv. 2021;Arxiv ID: 2110.00563.

32. Kojima M. Application of Multi-Armed Bandits to Model-assisted designs for Dose-Finding Clinical Trials. Arxiv. 2021;Arxiv ID: 2201.05268.

33. Lin R, Yin G, Shi H. Bayesian adaptive model selection design for optimal biological dose finding in phase I/II clinical trials. Biostatistics 2021.

34. Dykstra LR, Robertson T. An algorithm for isotonic regression for two or more independent variable. Ann Stat;10(3):708-716

**Table 1.** The number of patients treated for dose retainment (TTL=0.3)

| Design | Number of patients treated at current dose-combination | | | | | |
|---|---|---|---|---|---|---|
| | 3 | 6 | 9 | 12 | 15 | 18 |
| Keyboard | 1 | 2 | 3 | 4-5 | 4-5 | 5-6 |
| BOIN | 1 | 2 | 3 | 3-4 | 4-5 | 5-6 |

If the current dose-combination is the maximum dose-combination, the dose-combination retained even if it is below the number of patients treated above. If the current dose-combination is the minimum dose-combination, the dose-combination retained even if it is over the number of patients treated above.

**Table 2.** The number of patients treated and DLTs for each dose combination $(n_{j,k}, m_{j,k})$

| $d_{1,1}$ | (3,0) | $d_{1,2}$ |       | $d_{1,3}$ |       |
|---|---|---|---|---|---|
| $d_{2,1}$ | (6,1) | $\boldsymbol{d_{2,2}}$ | **(9,3)** | $d_{2,3}$ | (3,2) |
| $d_{3,1}$ |       | $d_{3,2}$ | (3,2) | $d_{3,3}$ |       |

Dark gray indicates not administered. Bold indicates that the safety evaluation has completed at the current dose.

**Table 3.** Decision table for early completion on BOIN and Keyboard design with sample size of 36

[BOIN]

| #Patients at current dose | #DLTs at current dose | #Remining patients |
|---|---|---|
| 3 | 1 | ≤2 |
| 6 | 2 | ≤2 |
| 9 | 3 | ≤12 |
| 12 | 3-4 | ≤24 |

Calculated using normal dose retainment probability adjusted by pool-adjacent-violators algorithm.

[Keyboard]

| #Patients at current dose | #DLTs at current dose | #Remining patients |
|---|---|---|
| 3 | 1 | ≤2 |
| 6 | 2 | ≤2 |
| 9 | 3 | ≤3 |
| 12 | 3-4 | ≤12 |

Calculated using normal dose retainment probability adjusted by pool-adjacent-violators algorithm.

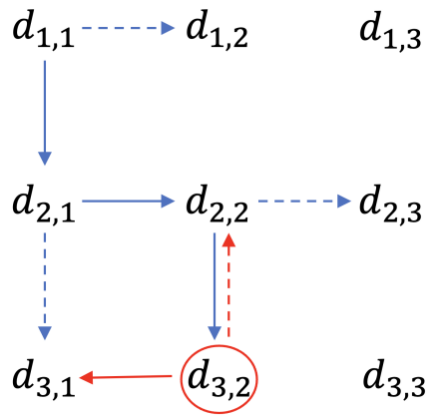

**Figure 1.** Example of strategy for dose-assignment

The solid line means that the drug was administered, and the dotted line means that the drug was a candidate for doses escalation/de-escalation. The blue means the dose escalation and the red means the dose de-escalation.

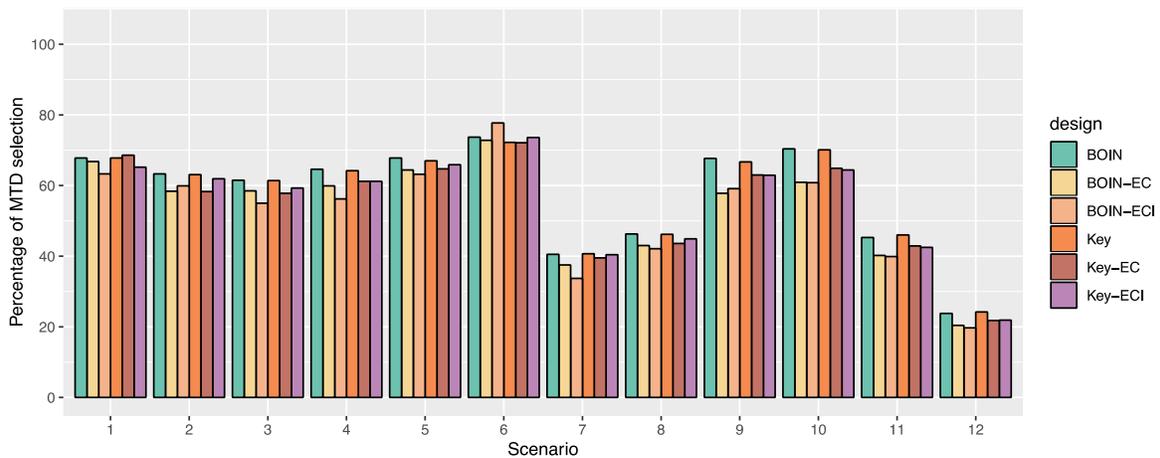

**Figure 2.** Percent change relative to the non-early completion version for selection of the

correct MTD

Key: Keyboard, EC: Early completion based on normal dose retainment probability, ECI: Early completion based on dose retainment probability using bivariate isotonic regression

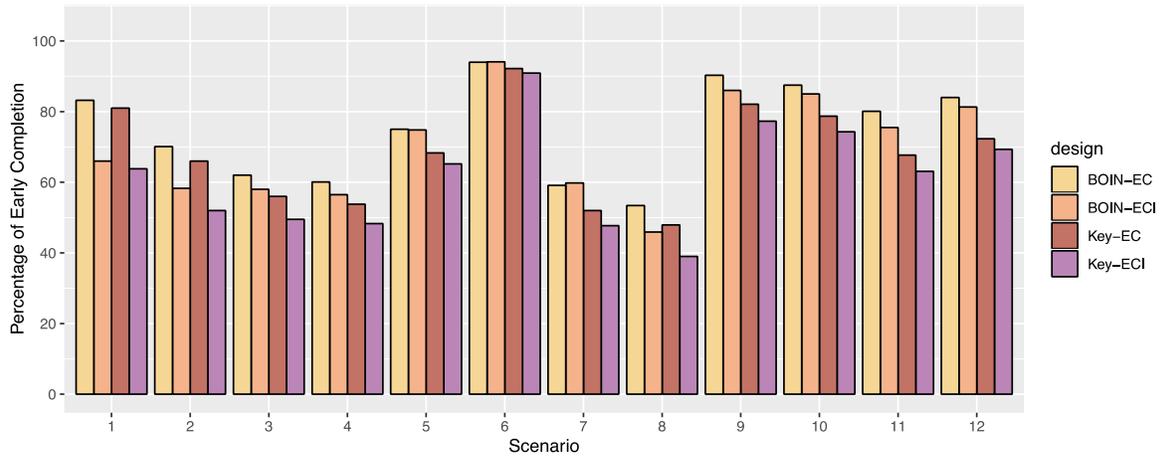

**Figure 3.** Percentage of early completion

Key: Keyboard, EC: Early completion based on normal dose retainment probability, ECI: Early completion based on dose retainment probability using bivariate isotonic regression